\documentclass[fleqn,11pt]{article}

\newcommand{\beq}{\begin{eqnarray}}

\newcommand{\eeq}{\end{eqnarray}}

\newcommand{\AmS}{{\protect\the\textfont2  
  A\kern-.1667em\lower.5ex\hbox{M}\kern-.125emS}}
\def\simge{\mathrel{%
   \rlap{\raise 0.511ex \hbox{$>$}}{\lower 0.511ex \hbox{$\sim$}}}}
\def\simle{\mathrel{
   \rlap{\raise 0.511ex \hbox{$<$}}{\lower 0.511ex \hbox{$\sim$}}}}

\newcommand{\BQ}{\begin{equation}}
\newcommand{\EQ}{\end{equation}}
\newcommand{\BQA}{\begin{eqnarray}}
\newcommand{\EQA}{\end{eqnarray}}
\newcommand{\be}{\begin{eqnarray}}
\newcommand{\ee}{\end{eqnarray}}
\newcommand{\NN}{\nonumber \\}
\newcommand{\del}{\partial}

\newcommand{\x}{x_\perp}
\newcommand{\y}{y_\perp}
\newcommand{\z}{z_\perp}
\newcommand{\rr}{r_\perp}
\newcommand{\q}{q_\perp}
\newcommand{\kk}{k_\perp}

\newcommand{\la}{\lambda}

\begin{document}

\begin{flushright}
SACLAY-T02/027   
\end{flushright}

\vspace{1cm}

\begin{center}
{\LARGE\bf Geometric Scaling above the Saturation Scale}\\

\vspace{0.9cm}

{\Large Edmond Iancu$^{\rm a}$, Kazunori Itakura$^{\rm a,b}$, 
and  Larry McLerran$^{\rm c}$}\\

\vspace{5mm}

{\large\it $^{\rm a}$~Service de Physique Theorique, CE Saclay, F-91191 
        Gif-sur-Yvette}

\vspace{0.1cm}

{\large\it $^{\rm b}$~RIKEN BNL Research Center, BNL, Upton NY 11793}

\vspace{0.1cm}

{\large\it $^{\rm c}$~Nuclear Theory Group, Brookhaven National Laboratory,
        Upton, NY 11793  } 

\vspace{0.5cm}

\end{center}

\vspace{0.5cm}

\begin{abstract}
We show that the evolution equations in QCD predict geometric scaling
for quark and gluon distribution functions
in a large kinematical window, which extends above
the saturation scale up to momenta $Q^2$ of order $100~{\rm GeV}^2$. 
For $Q^2 < Q^2_s$, with $Q_s$ the saturation momentum, 
this is the scaling predicted by the Colour Glass Condensate and
by phenomenological saturation models.
For $1 \simle \ln(Q^2/Q_s^2) \ll \ln(Q_s^2/\Lambda^2_{\rm QCD})$,
we show that the solution to the BFKL equation shows approximate scaling, 
with the scale set by $Q_s$. 
At larger $Q^2$, this solution does not scale any longer.
We argue that for the intermediate
values of $Q^2$ where we find scaling, 
the BFKL rather than the double logarithmic
approximation to the DGLAP equation properly describes the dynamics.
We consider both fixed and running couplings, with the scale
for running set by the saturation momentum. The anomalous dimension
which characterizes the approach of the gluon distribution function
towards saturation is found to be close to, but lower than, one half.
\end{abstract}

\newpage
\section{Introduction}
\setcounter{equation}{0}

At very high energies, the contribution to the hadronic wavefunction
which dominates hadronic processes corresponds to a state of very high
partonic (mostly gluonic) density. 
This high density matter is believed to reach a 
saturation regime, and become a Color Glass Condensate \cite{MV94,PI}.
(See also \cite{Cargese} for a recent review and more references.)
Recent phenomenological analysis of the data from HERA and the
RHIC  show results qualitatively and semi-quantitatively 
consistent with this picture \cite{Dima}.
The results, while not entirely compelling, provide a strong motivation for
the further theoretical investigation of this novel form of high 
density matter.

In a very important paper \cite{geometric}, 
Sta\'sto, Golec-Biernat and
Kwieci\'nski have shown that the HERA
data on deep inelastic scattering (DIS) at low $x$ \cite{z}, 
which are a priori functions of two independent variables
--- the photon virtuality 
$Q^2$ and the Bjorken variable $x$ ---, 
are consistent with scaling in terms of the variable
\beq\label{calT}
	{\cal T} = Q^2 R^2_0(x)
\eeq
where $R^2_0(x) = (x/x_0)^{\lambda}/Q_0^2$ with 
$\lambda = 0.3 - 0.4$,  $Q_0 = 1 ~{\rm GeV}$, and
$x_0 \sim 3 \times 10^{-4}$ in order to fit the data.
 In particular, the data 
for the virtual photon total cross section at $x < 0.01$ and 
$Q^2<400~{\rm GeV}^2$
are consistent with being only a function of ${\cal T}$. 
Note that, because
$
	\sigma_{\gamma^* p} = {4\pi^2}{\alpha_{\rm EM}} F_2(x,Q^2)/ Q^2,
$
this implies that $F_2(x,Q^2)/Q^2$ is a function only of ${\cal T}$
in the indicated kinematical range.

At first sight, it is tempting to interpret this scaling as a consequence
of the existence of the Color Glass Condensate: Indeed, 
it has been argued that in the saturated region
the various parton densities are functions only
of $Q^2/Q_s^2(x)$, with $Q_s$ the saturation momentum,
which is proportional to the gluon density, and thus
increases rapidly with $1/x$. 
Although $Q_s$ has not been fully determined from
first principles, its estimates based on
the quantum evolution of the  Color Glass Condensate suggest
indeed a power-like increase: $Q_s^2 \sim Q_0^2\, x^{-\lambda}$
\cite{AM2,Levin-Tuchin,SAT,AB01,Motyka}. (This will be further discussed
in this paper.) 
The fundamental problem with this argument is that it 
is valid only for $Q^2$ less
than or of the order of the saturation momentum, which is at most
several ${\rm GeV}^2$, while the fit of Ref. \cite{geometric}
extends up to $Q^2$ of the order of several hundred GeV$^2$.

The main purpose of this paper is to show that the scaling region 
for the various 
distribution functions is in fact much larger than the saturation region.
We shall find geometric scaling for all momenta $Q^2$ 
such that the following inequality is satisfied:
\beq\label{lnineq}
	\ln(Q^2/Q^2_s) \ll \ln(Q_s^2/\Lambda_{\rm QCD}^2).
\eeq
For $Q_s \sim 2~{\rm GeV}$ and $\Lambda_{\rm QCD} \sim
200~$MeV, the upper scale on $Q^2$ is $Q_s^4/\Lambda_{\rm QCD}^2
\sim 400~ {\rm GeV}^2$.
While at soft momenta $Q^2 \simle Q_s^2$ this scaling is an
expected consequence of saturation, at high momenta
$1 < \ln(Q^2/Q_s^2) \ll \ln(Q_s^2/\Lambda_{\rm QCD}^2)$ 
it rather corresponds
to a regime where parton densities are small, and linear evolution
equations apply.  And indeed we shall see that the {\it extended} scaling
at $Q^2 > Q_s^2$ arises from solutions to the BFKL equation \cite{BFKL},
which is the appropriate limit of the general non-linear evolution equations 
[2,13--18] in the kinematical range of interest.

Note that, in this paper, we shall use the BFKL equation as
an {\it effective} equation, valid in some range of $Q^2$ whose
increase in $Q^2$ is correlated with the decrease in Bjorken $x$
(because of the constraint $Q^2 > Q_s^2(x)\sim Q_0^2\, x^{-\lambda}$).
Thus, conceptual difficulties of BFKL, like the exponential
increase of its solution at high energy, and its ``infrared diffusion'',
will be of no concern for us. In fact, an essential step in our analysis
will be to clarify the kinematical ranges in which the solution to the
BFKL equation (in the saddle point approximation) is of the ``genuine
BFKL type'' (i.e., it is close to the saddle point which
governs the limit $x \rightarrow 0$ with $Q^2$ fixed), or of the
``double logarithmic type'' (i.e., close to the saddle point which describes
the limit $Q^2\to\infty$ with $x$ fixed).
As well known, in the latter limit the BFKL solution becomes equivalent
to the double logarithmic approximation (DLA) to the DGLAP equation \cite{DGLAP}.
But this limit turns out not to be relevant for the geometric scaling:
Indeed, the window for extended geometric scaling that we shall
find is actually in the range controlled by the BFKL saddle point.

It is quite remarkable that even at high momenta $Q^2 > Q_s^2$,
 it is still the saturation momentum $Q_s(x)$
which sets the scale in the scaling variable (i.e.,
which plays the role of $1/R_0(x)$ in eq.~(\ref{calT})).
This is so because the solution to the BFKL equation 
must properly match when $Q^2 \to Q_s^2$
on the corresponding solution at saturation. 

In brief, our strategy in this paper will be as follows:
We start with a solution to the BFKL equation, 
which in general (i.e., for arbitrary $Q^2$) does not appear to scale.
By extrapolating
this solution down to momenta $Q^2\sim Q_s^2(x)$, we shall estimate
the saturation scale $Q_s(x)$ from the matching condition with the
solution at saturation. The latter is not precisely known around $Q_s$,
but for the present purposes such a precise solution is not really needed:
All that we need is a {\it saturation criterion} which gives us the value
of the quantity of interest at $Q^2= Q_s^2(x)$. This criterion is
simply that the gluon distribution function itself becomes of
order $1/\alpha_s$. Such a criterion follows
from the general analysis of the non-linear effects due to high parton 
densities [1--3,7--11,13--18,20,21].
By expanding the BFKL solution around $Q^2=Q_s^2(x)$, we shall find geometric
scaling in the momentum range where the first term in
this expansion is a good approximation, namely for
 $1<  \ln(Q^2/Q_s^2) \ll \ln(Q_s^2/\Lambda_{\rm QCD}^2)$.
This conclusion holds both for a fixed coupling and for a running coupling
(with the scale for running set by the saturation momentum), although the
saturation scales turn out to be different in the two cases.

Note that, although the saturation is crucial for the arguments in this
paper, the emergence of an { extended} scaling window above the saturation 
scale is {\it not} an automatic consequence of the saturation. Rather, this
requires some non-trivial properties of the linear evolution equations as well.
To see this, consider some dimensionless functions $f(x,Q^2)$ (e.g., the
gluon distribution) which at high $Q^2$ obeys a linear evolution equation
(BFKL or DGLAP), but at $Q^2\sim Q^2_s(x)$ satisfies a saturation condition
of the type (after an appropriate normalization):
\be\label{GS}
f(x,Q^2=Q^2_s(x))\,=\,1.\ee 
As a consequence of this condition, and without loss of generality,
the function $f$ near $Q^2=Q^2_s(x)$ 
(i.e., for $|\ln(Q^2/Q_s^2)| \ll \ln(Q_s^2/\Lambda_{\rm QCD}^2)$),
can be approximated as:
\be\label{A}
f(x,Q^2)\,\simeq\,\left(\frac{Q^2}{Q^2_s(x)}\right)^{\lambda_s(x)}\,,\ee
where $\lambda_s(x)\equiv \big(\partial \ln f/\del\ln Q^2\big)|_{Q^2_s(x)}$.
Assume that this limited expansion can be extended up to values $Q^2$ which
are high enough for the non-linear effects to be subleading.
Then, the exponent $\lambda_s(x)$ can be determined by solving
the BFKL equation with the saturation boundary condition (\ref{GS}).
A priori, the result $\lambda_s(x)$ can be a
non-trivial function of $x$. If this happens, then  eq.~(\ref{A}) shows
no geometric scaling. However, in our subsequent analysis, we shall discover
that for solutions to the BFKL equation, $\lambda_s$ is
a number independent of $x$ --- as a consequence of the scale-invariance
of the BFKL kernel ---, so that $f(x,Q^2)$ shows scaling in the
window of validity of the expansion (\ref{A}). The difference between the
actual value of the exponent $\lambda_s$ and its ``na\"{\i}ve'' value
expected from perturbative considerations at large $Q^2$
will be referred to as the ``anomalous dimension''.

The saturation criterion that we shall actually
use refers to the scattering of a small
dipole off a hadronic target: $1/Q_s$ is the critical transverse size of 
the dipole at which the scattering amplitude becomes of order one
(``black disk limit'') \cite{AM0,SAT,K,GBW99,Levin-Tuchin}.
This criterion, which is equivalent (at least, for our present purposes) 
with the more standard  condition on the  gluon distribution 
\cite{MV94,AM2,SAT,GLR,MQ,JKMW97,KM98}, is simpler to use in
applications to deep inelastic scattering, since more directly related to the
corresponding cross-section at small $x$ \cite{AM0,AM2,NZ91,FS98} :
\be\label{sigmaDIS}
\sigma_{T,L}(\tau,Q^2)\,=\,
\int_0^1 dz \int d^2r_\perp\,|\Psi_{T,L}(z,r_\perp;Q^2)|^2\,
\hat\sigma(\tau,r_\perp).\ee
In this equation, $\sigma_{T,L}(\tau,Q^2)$ is the cross-section for the scattering 
of a virtual transverse ($T$) or longitudinal ($L$) photon off the hadron,
at relative rapidity $\tau\equiv \ln(1/x)\sim \ln s$.
(Note that, from now on, we use $\tau$ to indicate the
dependence upon the total center-of-mass energy squared $s$.) Furthermore, $Q^2$ is
(minus) the photon virtuality, and $\Psi_{T,L}(z,r_\perp;Q^2)$ is the 
light-cone wavefunction for the photon splitting 
into a $q\bar q$ pair  (the ``dipole'') with transverse size $r_\perp$ and
a fraction $z$ of the photon's longitudinal momentum carried
by the quark.
Finally, $\hat\sigma(\tau,r_\perp)$ is the dipole-hadron
cross-section, which can be computed in the eikonal approximation as:
\be\label{sigmadipole}
\hat\sigma(\tau,r_\perp)\,=\,2\int d^2b_\perp\,
(1-S_\tau(x_{\perp},y_{\perp})),\qquad 
S_\tau(x_{\perp},y_{\perp})\equiv \frac{1}{N_c}\,
\langle {\rm tr}\big(V^\dagger(x_{\perp}) V(y_{\perp})\big)
\rangle_{\tau},\ee
with $r_\perp=x_{\perp}-y_{\perp}$ (the
quark is at $x_{\perp}$, and the antiquark at $y_{\perp}$), and
the impact parameter $b_\perp=(x_{\perp}+y_{\perp})/2$.
In eq.~(\ref{sigmadipole}), $V^\dag$ (or $V$) is a Wilson line along the straight 
line trajectory of the quark (or the antiquark), that is, 
a path ordered exponential of the color field created 
in the hadron by color source at rapidities $\tau'<\tau$.
The brackets in the definition of the $S$-matrix element 
$S_\tau(x_{\perp},y_{\perp})$ refer to the average over all configurations
of these colour sources \cite{MV94}.

A small dipole is weakly interacting:
$S_\tau(r_{\perp})\approx 1$ for $\rr \ll 1/Q_s(\tau)$.
A large dipole, on the other hand, is strongly absorbed: 
$S_\tau(r_{\perp})\ll 1$ for $\rr \simge 1/Q_s(\tau)$.
This is so because of the large density of the saturated gluons
in the hadron wavefunction. 
At momenta $Q^2\simle Q_s^2(\tau)$, there is only one intrinsic
scale in the problem, the saturation momentum $Q_s(\tau)$ itself 
(we mean this for a fixed impact parameter and a fixed coupling).
So, all physical quantities
should be expressed as a dimensionless function of $Q^2/Q_s^2$ times
some  power of $Q_s^2$ giving the right dimension.
These general properties expected for $\hat\sigma(\tau,r_\perp)$
have been incorporated in a simple ``saturation'' model by
Golec-Biernat and W\"usthoff \cite{GBW99} 
($1/R^2_0(x)$ plays the role of the saturation scale) :
\be\label{GBDIP}
\hat\sigma(\tau,r_\perp)\,=\,\sigma_0\Big(1-{\rm e}^{-\rr^2/4R_0^2(x)}\Big)
\,\equiv\,\sigma_0 \,g\left(\frac{\rr^2}{4R^2_0(x)}\right),\ee
which appears to give a reasonable description of the HERA data at small $x$.
By inserting this expression in eq.~(\ref{sigmaDIS}), one can check 
that the scaling property is transmitted to the DIS cross-section:
$\sigma_{\gamma^* p}(\tau,Q^2)=\sigma_{\gamma^* p}({\cal T})$
with ${\cal T}$ of eq.~(\ref{calT}) \cite{geometric}.

Written as it stands, eq.~(\ref{GBDIP}) shows exact
scaling for all distances $\rr$, and not only in the saturation regime
$\rr\gg R_0(x)$. In reality, however, we know this scaling to be
violated at sufficiently small distances $\rr\ll R_0(x)$, where
eq.~(\ref{GBDIP}) should be replaced by \cite{AM0,LR87,AGL97}
\be\label{GM-DIP}
\hat\sigma(\tau,r_\perp)\,=\,\sigma_0\left(1-{\rm exp}\bigg\{-\rr^2
\frac{\pi^2\alpha_s}{N_c}\frac{xG(x,1/\rr^2)}{\sigma_0}\bigg\}\right),\ee
where $xG(x,1/\rr^2)$ is the gluon distribution function evaluated
at $Q^2=1/\rr^2\gg Q_s^2$. At such high momenta, $xG(x,Q^2)$ is a
solution to some linear evolution equation, usually taken to be the 
DGLAP equation, which, at least for sufficiently large
$Q^2$, has no scaling behaviour (see Sect. 3 below). It will be our main
objective in what follows to establish up to which momenta $Q^2$
the scaling property holds for the dipole cross-section, and thus for
the quark distribution in the hadron. 
To characterize this property also for the gluon distribution,
we shall find it convenient to introduce a definition of the latter
in terms of the dipole scattering amplitude (see eq.~(\ref{GM-DIP})
and eqs.~(\ref{G-def})--(\ref{glue}) below).

To simplify the problem, we shall assume 
that the dependence upon the 
impact parameter is not essential for the present purposes,
so we can treat the  hadron as being homogeneous 
in the transverse plane\footnote{This assumption 
is implicit in eqs.~(\ref{GBDIP}) and (\ref{GM-DIP}),
where $\sigma_0=2\int d^2b_\perp=2\pi R^2$, with $R$ the hadron
radius.}.
This implies 
$S_\tau(x_{\perp},y_{\perp})=S_\tau(\rr)$, with $\rr$ the size of
the dipole, and the scaling properties
of $\sigma_{\gamma^* p}(\tau,Q^2)$ are directly related to corresponding
properties of $S_\tau(\rr)$, that we shall study.

The function $S_\tau(\rr)$ will be obtained by solving an
appropriate evolution equation in $\tau$. As already mentioned,
we shall be mainly concerned with solutions to 
the BFKL equation, but 
this equation will be viewed as the linear limit of more general,
non-linear, evolution equations. This is important, as it will allow
us to keep trace of the non-linear effects via the boundary
condition at $Q^2\sim Q_s^2(\tau)$ (the ``saturation condition'').

It is possible to write down a formal evolution 
equation for $S_\tau(x_{\perp},y_{\perp})$ without specifying
the average on the target hadron in the definition (\ref{sigmadipole})
\cite{B,K,Braun}.
In general, such an equation is only the starting point of a hierarchy of
coupled evolution equations for the correlators of Wilson line operators
\cite{B}, which can be encoded in a single, functional evolution equation
\cite{W}. But a closed equation for 
$S_\tau(x_{\perp},y_{\perp})$ can still be obtained in the
large $N_c$ limit. This has been derived by Kovchegov \cite{K} within
the Mueller's dipole model \cite{AM3}, and follows also
from the general evolution equations by Balitsky \cite{B} by taking
the large $N_c$ limit. (See also \cite{Braun} for a different derivation.)
We shall refer to this as the Balitsky-Kovchegov (BK) equation. 

Alternatively, an effective theory  can be constructed
for the small-$x$ component of the hadron wavefunction
\cite{MV94,PI,Cargese,SAT,JKMW97,K96,JKLW97}, thus
allowing a direct and explicit calculation of the average
in eq.~(\ref{sigmadipole}).
In this effective theory, the high density gluon configurations at small $x$
are treated as a ``Colour Glass'', i.e., as the classical
colour field generated by colour sources at rapidities
 $\tau'< \tau = \ln(1/x)$, which
are ``frozen'' in some random configuration. In this picture the
evolution is viewed as a renormalization group operation in which layers
of quantum fluctuations are successively integrated out to generate
the sources \cite{JKLW97,PI}. This leads to a functional
renormalization group equation (RGE) for
the probability distribution of the colour sources.
Remarkably, this RGE
is equivalent  \cite{PI} to the Wilson line approach 
of Refs. \cite{B,K,W}:
it generates the same evolution equations for the correlation
functions of Wilson lines.
At high transverse momenta $Q^2$, the non-linear effects are weak
and can be expanded out. To lowest order in this expansion,
the general RGE (and also the BK equation)
reduces to the BFKL equation \cite{BFKL}, as already mentioned.

The outline of this paper is as follows:

In the second section, we rely on the BK equation to 
explore the scaling properties of
distribution functions in the deeply saturated region where the
gluon density is of order $1/\alpha_s$. {By assuming that scaling solutions
exist,} we determine the dependence of the saturation momentum
upon $\tau$ for both the fixed coupling and the running coupling cases.
Throughout this paper, the scale
for the running of $\alpha_s$ is always the saturation momentum.

In the third section, we consider the intermediate $Q^2$ region
where the evolution equation linearizes and reduces to the BFKL
equation. We construct approximate solutions for this equation
in the saddle point approximation and show that the solution
which is relevant for the approach to saturation and the extended
scaling is closer to the standard BFKL solution, rather than to
the double-logarithmic DGLAP-like solution.  We determine the 
saturation momentum, compute the anomalous dimension of the distribution
function, and show the solution has geometric scaling for both
the cases of running and fixed coupling. 

The last section contains a summary of our results, and some discussion.

\section{Scaling properties of the Balitsky-Kovchegov equation}
\setcounter{equation}{0}

In this section, we shall briefly discuss
 the BK equation \cite{B,K},
which is a non-linear equation for the evolution of the dipole-hadron scattering 
amplitude with $\tau=\ln(1/x)$. This equation will be used here
just for qualitative arguments, which by themselves are valid for
any $N_c$, although the BK  equation holds, strictly speaking, only in the large
$N_c$ limit. (The same arguments at finite $N_c$
would require the full formalism in Refs. \cite{B,W,PI,Cargese}.)
Specifically, the following properties of the BK equation will
be useful for what follows: First, it covers both the linear and the
non-linear regimes of the quantum evolution (according to the value of $Q^2$), 
second, it reduces to the BFKL (and eventually DLA) equation
in the linear regime at high $Q^2$, and, third, 
it demonstrates the relation between geometric scaling and saturation
at low $Q^2$. 

In fact, we shall find that the structure of the BK equation is consistent
with scaling solutions, although the kind of global arguments
that we shall use cannot decide if such solutions actually exist or not.
This will open the discussion of scaling solutions outside the saturation regime,
to be pursued in the framework of linear evolution equations in Sect. 3.
The scaling properties of the Balitsky-Kovchegov equation have been
also investigated numerically, in Refs. \cite{Levin-Tuchin,Motyka,L01}.

The BK equation is most succinctly written as an equation for
the $S$-matrix element $S_\tau(\rr)$
for dipole-hadron scattering, eq.~(\ref{sigmadipole})
(below,  $\bar \alpha_s= N_c\alpha_s/\pi$) :
\beq
	{\partial \over {\partial \tau}} S_\tau( x_\perp-\y) & = &
-\bar \alpha_s \int {{d^2z_\perp} \over {2\pi}} 
{{(x_\perp-\y)^2} \over {(x_\perp-z_\perp)^2 (\y-z_\perp)^2}} 
\nonumber \\ & &
\times \Big(S_\tau(x_\perp-\y)- S_\tau(x_\perp-z_\perp)
S_\tau(z_\perp-\y)   \Big).\label{BK}
\eeq
Given $S_\tau(\rr)$, the
dipole-hadron scattering amplitude ${\cal N}_\tau(\rr)$ is obtained as:
\beq
	{\cal N}_\tau(\rr)\, = \,1 - S_\tau( \rr).
\eeq
For a small dipole, $\rr\ll 1/Q_s(\tau)$,  ${\cal N}_\tau(\rr)$ is small
as well, ${\cal N}_\tau(\rr)\ll 1$, and one can linearize eq.~(\ref{BK})
with respect to ${\cal N}_\tau(\rr)$: one then obtains the BFKL equation.  
This can be easily seen in the momentum space form of the BK equation.
(The BFKL equation in coordinate space will be discussed
in the next section.) Specifically, if one defines $\varphi_\tau(\kk)$ by 
\BQ
\frac{{\cal N}_\tau(\rr)}{\rr^2}=\int \frac{d^2\kk}{(2\pi)^2}
\,{\rm e}^{i\kk\cdot \rr} \,\frac{\varphi_\tau(\kk)}{\kk^2},
\label{Fourier}
\EQ
one finds 
\BQ
\frac{\del}{\del\tau}\varphi_\tau(\q)=\bar\alpha_s
\int {d^2 k_{\perp}\over \pi}\,
 {q^2_{\perp} \over k^2_{\perp} (q_{\perp}-k_{\perp})^2}\,
\left(\varphi_\tau(\kk) -\,{1\over 2}\,\varphi_\tau(q_{\perp})\right)
-\frac{\bar\alpha_s}{2\pi} \frac{(\varphi_\tau(\q))^2}{\q^2} \,.
\label{BK_mom}
\EQ
The small-dipole condition ${\cal N}_\tau(\rr)\ll 1$ in coordinate space
corresponds to the condition $\varphi_\tau(\q)/\q^2\ll 1$ in momentum space,
which is satisfied provided $\q^2$ is large enough, $\q^2\gg Q_s^2(\tau)$.
In this regime, the last term, quadratic in $\varphi_\tau(\q)/\q^2$, can be neglected
in the r.h.s. of eq.~(\ref{BK_mom}), which then reduces to the BFKL equation
\cite{BFKL}. For momenta $\q^2$ which are even larger, this becomes
equivalent to the double-log limit of the DGLAP equation (see Sect. 3 below).

On the other hand, non-linear effects become crucial at low momenta 
$\q^2 \simle Q_s^2(\tau)$, that is, for a large dipole which is strongly absorbed 
by the hadron: $S_\tau(\rr)\ll 1$ for $\rr \gg 1/Q_s(\tau)$. 
In this regime, the right hand side of eq.~(\ref{BK}) can be simplified by
neglecting the term quadratic in $S_\tau\,$; this is appropriate since the
dominant contribution (in the sense of the leading log) comes from
$\z$ satisfying $1/Q_s(\tau)\ll |\z-\x|\ll \rr$ (or a similar condition
on $|\z-\y|$). The equation then reduces to
\beq\label{S-low}
	{\partial \over {\partial \tau}} S_\tau(\rr) \,\simeq\,
-\bar \alpha_s \ln\Big(\rr^2 Q_s^2(\tau)\Big)\,S_\tau(\rr),\ee
for which a solution will be written down shortly. (This requires the $\tau$
dependence of the saturation scale $Q_s^2(\tau)$.)

Still in the non-linear regime, it is interesting to determine
the limiting form of the function $\varphi_\tau(\q)$ at low $\q$. By solving
eq.~(\ref{BK_mom}) in this regime, or, more directly,
by noticing that ${\cal N}_\tau(\rr)\simeq 1$
for $\rr > 1/Q_s(\tau)$ in eq.~(\ref{Fourier}), one deduces that:
\BQ
\varphi_\tau(\q^2 \ll Q_s^2)\,\simeq\, \q^2\int_{1/Q_s^2}^{1/\q^2}\frac
{d^2\rr}{\rr^2}\,
= \,\pi \q^2 \ln \frac{Q_s^2(\tau)}{\q^2}.
\label{asymptotic}
\EQ
It can be checked that this is indeed a solution to eq.~(\ref{BK_mom})
at low $\q^2$ and to leading-log accuracy \cite{MFA}.

At this stage, one may notice that the properties of the function
$\varphi_\tau(\q)$ introduced in eq.~(\ref{Fourier}) are very similar
to those expected for the ``unintegrated gluon distribution'', i.e.,
the density of gluons in the transverse phase-space. Indeed, at large momenta
$\varphi_\tau(\q)$ satisfies the BFKL equation and is truly proportional to the
gluon distribution, as manifest on eq.~(\ref{GM-DIP}). At low momenta,
$\q^2 \ll Q_s^2$, it has the same behaviour in $\q$,
eq.~(\ref{asymptotic}), as the gluon density
in the saturation regime \cite{AM2,SAT,Cargese}. It is thus quite 
reasonable to define the gluon phase space distribution function via
the Fourier transform of the dipole scattering amplitude (with 
$C_F=(N_c^2-1)/2N_c$):
\be\label{G-def}
(2\pi)^2\frac{d^5 N}{d\tau d^2k_\perp d^2 b_\perp}\,\equiv\,\frac{N_c^2-1}{
\pi^2\alpha_s C_F}\,\frac{\varphi_\tau(\kk,b_\perp)}{\kk^2},\ee
where we have reintroduced the impact parameter dependence, for more
generality (so, \\ ${\varphi_\tau(\kk,b_\perp)}/{\kk^2}$ is actually
the Wigner transform of ${\cal N}_\tau(\rr,b_\perp)/\rr^2$), and 
the overall normalization follows by comparison with
eq.~(\ref{GM-DIP}).
This gives a gluon distribution:
\beq
xG(x, Q^2) \equiv \frac{d N}{d\tau} \,=\,  {{2N_c} \over {\pi^2 \alpha_s}} 
\int^{Q^2} {{d^2k_\perp } \over {(2\pi)^2}} 
\int d^2b_\perp\int d^2r_\perp  \,{\rm e}^{-ik_\perp \cdot r_\perp}
{{{\cal N}_\tau(r_\perp,b_\perp)}
\over r_\perp^2}\, .
\label{glue}
\eeq
The definition (\ref{G-def}) should be compared to the canonical definition
of the gluon density, which involves the expectation value
of the gluon occupation number in the infinite momentum frame and in
the light-cone gauge \cite{AM2,PI,Cargese}. In the linear regime at high $Q^2$, 
these two definitions are equivalent,
and provide both the standard gluon distribution which measures 
(Bjorken) scaling violation in $F_2$. But in the non-linear regime
at $Q^2\simle Q^2_s$, there is no simple relationship between $F_2$
and the canonical gluon density. By contrast, the definition (\ref{G-def}) 
has the advantage that it appears linearly in the expression for $F_2$,
so it is related to a directly measurable quantity even at saturation.
In particular, all the scaling properties that we shall establish later
for the dipole  scattering amplitude ${\cal N}_\tau(\rr)$ will immediately
translate to the gluon distribution defined as in eqs.~(\ref{G-def})--(\ref{glue}),
and we expect them to hold for the canonical gluon distribution as well.

Note the factor of $1/\alpha_s$ in the relation  (\ref{glue}) between 
${\cal N}_\tau(r_\perp)$ and $xG(x, Q^2)$. 
Via the running of $\alpha_s$, 
this will be a source of geometric scaling violation in $xG(x, Q^2)$, 
as we shall show later that ${\cal N}_\tau(r_\perp)$ shows scaling in the range
specified by eq.~(\ref{lnineq}).

Eqs.~(\ref{S-low}) and (\ref{asymptotic}) confirm already our expectation that,
in the saturation regime, the dipole scattering amplitude 
should exhibit geometric scaling: the solution $S_\tau(\rr)$ to eq.~(\ref{S-low})
is clearly a function of $\rr^2 Q_s^2(\tau)$ alone, and, similarly,
 the dimensionless quantity
$\varphi_\tau(\q)/\q^2$ is only a function of $\q^2/Q_s^2(\tau)$
when $\q^2\ll Q_s^2(\tau)$. Encouraged by this observation, 
let us investigate the scaling properties of the solution to the BK equation
in more generality:

\bigskip
\noindent{\it 1) Fixed coupling case:} 

We consider first the case where the coupling $\alpha_s$ in the r.h.s.
of eq.~(\ref{BK}) is fixed, and search for a solution $ S_\tau(\rr)$ to 
this equation in the scaling form:
\beq
	S_\tau(\rr) \equiv 1-\Phi(\xi),\quad \xi\equiv
\ln \frac{1}{\rr^2 Q_s^2(\tau)}.\label{scaling_assumption}
\eeq
Assuming such a scaling, the $\tau$ dependence of the saturation scale 
is then fixed by the equation. To see this, 
note first that for any function $f(\xi)$, 
\BQA
\frac{\del}{\del\tau}f(\xi)=-\left(\frac{\del}{\del\tau}\ln Q_s^2(\tau)\right)
f'(\xi), \quad \rr^2 \frac{\del}{\del\rr^2}f(\xi)=-f'(\xi),\nonumber
\EQA
with $f'(\xi)=df(\xi)/d\xi$.
Then, integrating the BK equation (\ref{BK}) 
over $r_\perp=\x-\y$ (after first dividing it by $r_\perp^2$), 
one gets 
\BQA
\int d^2r_\perp \frac{1}{r_\perp^2}\frac{\del}{\del \tau}S_\tau(r_\perp)= 
\pi \Big(S_\tau(\infty)-S_\tau(0)\Big)
\frac{\del}{\del \tau}\ln Q_s^2(\tau)=-\pi \frac{\del}{\del \tau}\ln Q_s^2(\tau),\nonumber
\EQA
where we have used the boundary conditions $S_\tau(0)=1$ and 
$S_\tau(\infty)=0$.
Therefore,
\BQ
\frac{\del}{\del \tau}\ln Q_s^2(\tau) = c\, \bar\alpha_s, \label{dif_eq}
\EQ
where $c$ is given by 
\BQ
c\equiv \int \frac{d^2r_\perp d^2\z}{2\pi^2}\,
\frac{1}{\z^2(r_\perp-\z)^2}\,\Big(S_\tau(r_\perp) - S_\tau(\z)S_\tau(r_\perp-\z)\Big).\label{c}
\EQ
If $S_\tau(\rr)$ is  a scaling solution, 
then the r.h.s. of eq.~(\ref{c}) is a constant independent of $\tau$. 
This follows from the scale invariance of the integrand: The function
(\ref{scaling_assumption}) depends upon $\tau$ only via the scale
$Q_s(\tau)$ within the scaling variable; thus,  by changing
variables according to $u_\perp^i\equiv r_\perp^i Q_s(\tau)$
and $v_\perp^i\equiv z_\perp^i Q_s(\tau)$,
all the $\tau$ dependence goes away.
More explicitly, the integral in eq.~(\ref{c})
can be written only in terms of the scaling variable 
$\xi=\ln 1/\rr^2Q_s^2(\tau)$: 
\BQ\label{c-BK}
c=\int_{-\infty}^\infty d\xi\int_{-\infty}^\infty d\xi'\,
{\rm e}^{-\xi}
\left[\frac{\Phi(\xi')-\Phi(\xi)}{|{\rm e}^{-\xi'}-{\rm e}^{-\xi}|}
+\frac{\Phi(\xi)}{\sqrt{4{\rm e}^{-2\xi'}+{\rm e}^{-2\xi}}}
\right]  -\frac{1}{2}\left(\int_{-\infty}^\infty d\xi\,\Phi(\xi)\right)^2.
\EQ
Therefore, the r.h.s. of eq.~(\ref{dif_eq}) is independent of $\tau$, 
which implies that the saturation scale grows 
exponentially with $\tau$ :
\BQ
Q_s^2(\tau)= \Lambda^2 \,{\rm e}^{c\bar\alpha_s \tau},
\label{saturation_scale_fixed}
\EQ 
with $\Lambda$ fixed by the initial condition (typically, 
$\Lambda\sim \Lambda_{\rm QCD}$).
With this saturation scale,  eq.~(\ref{S-low}) can be easily integrated, 
with the following result which shows how the black disk limit
($S_\tau=0$) is approached when $\rr\gg 1/Q_s$
\cite{Levin-Tuchin,SAT,Cargese}:
\beq\label{Ssat}
	S_\tau(\rr) \propto \exp\left\{ -{\xi^2 \over {2c}}
\right\}\,.
\eeq
In order to search for scaling solutions, it is preferable to
rewrite the original BK equation as an equation for the scaling function
$\Phi(\xi)$. It turns out that the equation takes a simpler form
when written in momentum space. That is, the scaling Ansatz 
(\ref{scaling_assumption}) is reformulated at the level of the BK
equation in momentum space (\ref{BK_mom}). In view of
eq.~(\ref{Fourier}), it is natural to write:
\BQ\label{psi}
\Psi(\zeta)\equiv \frac{\varphi_\tau(\q)}{\q^2}, \qquad \zeta \equiv 
\ln \frac{\q^2}{Q_s^2(\tau)}.
\EQ
The function $\Psi(\zeta)$ is related to $\Phi(\xi)$ via the following 
relation, which follows from eq.~(\ref{Fourier}):
\BQ\label{Fmom}
\Phi(\xi)= \int_{-\infty}^\infty \frac{d\zeta}{4\pi} 
\,{\rm e}^{\zeta-\xi}\, 
J_0\left({\rm e}^{(\zeta-\xi)/2}\right)
\Psi(\zeta)\, ,
\nonumber
\EQ
where $J_0$ is a Bessel function. By inserting the Ansatz (\ref{psi})
in eq.~(\ref{BK_mom}), and using also  eq.~(\ref{dif_eq}),
one obtains the following
equation for the scaling function $\Psi(\zeta)$:
\BQ\label{eqPsi}
-c \frac{\del}{\del \zeta}\Psi(\zeta)=
\int_{-\infty}^\infty d\zeta' 
\left\{
\frac{{\rm e}^{\zeta'}\Psi(\zeta')-{\rm e}^{\zeta}\Psi(\zeta)}
{|{\rm e}^{\zeta'}-{\rm e}^{\zeta}|}
+\frac{{\rm e}^{\zeta}\Psi(\zeta)}{\sqrt{4{\rm e}^{2\zeta'}+{\rm e}^{2\zeta}}}
\right\}-\frac{1}{2\pi}\Psi^2(\zeta)\, .
\EQ
This equation, together with the definition (\ref{c-BK}) of $c$,
and the relation (\ref{Fmom}) between $\Psi(\zeta)$ and $\Phi(\xi)$,
form a system of a coupled equations which in principle determine
the scaling solution and the associated coefficient $c$.

\bigskip 

\noindent{\it 2) Running coupling case:} 

One should stress that  the running of the QCD coupling is
a higher order effect, which so far has not been included in the derivation
of the non-linear evolution equations from first principles.
Thus, treating $\alpha_s$ in the BK equation (\ref{BK}) as a running coupling 
``constant'' is just a phenomenological way to incorporate some (potentially
large) higher order corrections, and suffers from ambiguities.
Here, we shall assume the one-loop--like running
\BQ
\bar\alpha_s(Q^2) 
= \frac{b_0}{ \ln(Q^2/\Lambda^2_{\rm QCD})} , \qquad 
b_0= \frac{12 N_c}{11N_c-2N_f}\,,
\label{running}
\EQ
with the scale $Q^2$ chosen as the saturation momentum: $Q^2=Q_s^2(\tau)$.
This is physically acceptable since $Q_s$ is the typical momentum of the
gluons in the saturated regime. Moreover, numerical 
solutions to the fixed coupling BK equation show that the dipole scattering 
amplitude is peaked around the saturation scale \cite{Motyka}. 
(Other possibilities for running, e.g., $Q^2=1/\rr^2$, with $\rr$ the size
of the dipole, will be left for a later study \cite{MFA}.)

Since the running coupling (\ref{running})
involves also the QCD scale $\Lambda_{\rm QCD}$, 
this is not a single scale problem any longer. 
It is therefore a nontrivial question in general whether a scaling solution 
exists with running coupling. But at least for very low $\q^2$,
the solution $\varphi_\tau(\q)$ in eq.~(\ref{asymptotic}) is manifestly a
scaling solution even for a running coupling, since independent of 
$\alpha_s$~! (Together with  eq.~(\ref{G-def}),
this reflects the fact that the gluon
density at saturation is of order $1/\alpha_s$.)

Thus, even in this case, it is legitimate to look for a scaling solution,
of the form (\ref{scaling_assumption}). By the same arguments as above,
this leads us to the differential equation 
\BQ\label{dif_eq_run}
\frac{\del}{\del \tau}\ln Q_s^2(\tau) = c\, \bar\alpha_s(Q_s^2(\tau)),
\EQ
where $c$ is again constant, and takes the same value 
as in the fixed coupling case, since determined (together
with the scaling function $\Phi(\xi)$)
by the same coupled equations (\ref{c-BK}), (\ref{Fmom}) and
(\ref{eqPsi}). As compared to the fixed coupling case, 
the growth of the saturation scale becomes somewhat milder
($\tau_0$ is an arbitrary constant)
\BQ\label{saturation_scale_run}
Q_s^2(\tau)=\Lambda_{\rm QCD}^2\ {\rm e}^{\sqrt{2 b_0 c 
(\tau +\tau_0)}}.
\EQ

It would be interesting to clarify if the system of equations 
(\ref{c-BK}), (\ref{Fmom}) and (\ref{eqPsi}) has indeed solutions,
i.e., if exact scaling solutions to the BK equation really exist.
But as we shall argue in the next section, the physically relevant solutions
necessarily violate scaling at sufficiently small $\rr$. This is so
since, when $\rr\ll 1/Q_s(\tau)$, the BK equation linearizes, so its
solution should match on the corresponding solution to the BFKL
(or DLA) equation, which shows scaling only in a limited range of
$\rr$ below the saturation length $1/Q_s(\tau)$.

\section{The BFKL equation in the context of saturation}
\setcounter{equation}{0}

We have seen in the previous section that, when the size of the 
dipole is small compared to the saturation scale, $\rr\ll 1/Q_s(\tau)$,
the general evolution equation can be linearized in the 
scattering amplitude ${\cal N}_\tau(\rr)$,  and then it reduces
to the (coordinate space form of the)  BFKL equation \cite{BFKL,AM3} :
\BQ\label{linearBK}
\frac{\del }{\del \tau} {\cal N}_\tau(\rr)=\bar\alpha_s \int 
{d^2z_\perp\over \pi}\, \, 
\frac{\rr^2}{(\rr-\z)^2\z^2}\, \left({\cal N}_\tau(\z)
-\frac12 {\cal N}_\tau(\rr)\right).
\EQ
The solutions to this equation have been extensively studied
(see, e.g., \cite{AKM} for an approach similar to ours), 
but they will be reconsidered here in the context of saturation, which requires
the function ${\cal N}_\tau(\rr)$ to satisfy the
boundary condition ${\cal N}_\tau(\rr)\sim 1$ for $\rr \sim 1/Q_s(\tau)$.
This is automatically satisfied by the solution to the non-linear BK equation,
but in the framework of the linear BFKL equation it becomes a non-trivial
condition which determines the saturation scale.
(A similar strategy to determine $Q_s$ in the context of the BFKL
equation has been previously used by A. Mueller \cite{AM2}.)

To understand this boundary condition, recall that, at saturation,
${\cal N}_\tau(\rr)=1-S_\tau(\rr)$ is a scaling function
(see, e.g., eq.~(\ref{Ssat})): ${\cal N}_\tau(\rr)=f(\rr^2Q_s^2(\tau))$,
and thus becomes a constant $\kappa$ 
when $\rr Q_s(\tau)=1$. This constant is a number of order one (although strictly
smaller: $\kappa < 1$),  since $S_\tau(\rr)\ll 1$ at saturation. The precise
value of $\kappa$ is a matter of convention --- it
defines what we mean exactly by ``the saturation scale'' ---,
but this will not matter for what follows.

\subsection{The solution to the BFKL equation revisited}

In this subsection we shall construct approximate solutions to the
BFKL equation (\ref{linearBK}) for the case of a fixed coupling constant $\alpha_s$.
Although some of the results are quite standard (see, e.g., \cite{AKM}), 
we prefer to go through their
derivation in some detail, in order to clarify the range of validity of the
various approximations. This will be important for the discussion of extended
scaling in the next subsection. Moreover, the techniques that we
introduce here will be also useful later.

The approximations that we shall perform rely on the following inequalities:
\be\label{INEQ}
\bar\alpha_s \tau\,\gg\,1,\qquad{\rm and}\qquad 
r=\ln \frac{Q^2}{\Lambda^2}\,\gg\,1,
\ee
where $\Lambda$ is some reference scale of order $\Lambda_{\rm QCD}$,
and $Q^2\equiv 1/\rr^2$.
The conditions (\ref{INEQ}) express the fact that we consider a perturbative
regime at small $x$ and large $Q^2$. In addition, the quantities
$r$ and $\bar\alpha_s \tau$ are not free to vary independently; rather, they
are constrained by $Q^2\gg Q_s^2(\tau)$ (which ensures that we are in
a linear regime), which requires (cf. eq.~(\ref{saturation_scale_fixed})):
\be\label{INEQ1}
r=\ln \frac{Q^2}{\Lambda^2}\,>\,\ln \frac{Q_s^2(\tau)}{\Lambda^2}\,=\,
c\bar\alpha_s \tau\,,\ee
with the coefficient $c$ to be determined in Sect. 3.2.

Note first that eq.~(\ref{linearBK}) has the same structure in coordinate space as
the usual BFKL equation in momentum space
(i.e., the linear part of eq.~(\ref{BK_mom})), 
so it can be solved via similar techniques. 
For the present purposes, it is convenient to use
the Mellin transform with respect to the transverse coordinates: 
\BQ\label{Mellin}
{\cal N}_\tau(\rr)=\int_{C} \frac{d\lambda}{2\pi i} 
\left(\frac{\rr^2}{\ell^2}\right)^\lambda  \chi_\tau(\lambda)
\EQ
where $\ell^2=1/\Lambda^2$ and  $\rr^2\ll \ell^2$, so that
the contour is taken on the {left} of all the singularities of the integrand
in the half plane Re $\lambda>0$.
 Since the BFKL equation is invariant under scale
transformations, the ensuing equation for $\chi_\tau(\la)$ is local in $\la$ : 
\BQ
\frac{\del }{\del \tau}\chi_\tau(\la)=
\bar\alpha_s \Big\{2\psi(1)-\psi(\lambda)-\psi(1-\lambda)\Big\}
\chi_\tau(\la),
\EQ
($\psi(\lambda)$ is the di-gamma function), and has the obvious solution
\BQ\label{solchi}
\chi_\tau(\lambda)\,=\,
{\rm e}^{\bar\alpha_s\tau\left\{2\psi(1)-\psi(\lambda)-\psi(1-\lambda)\right\}}\,
\chi_0(\lambda).\EQ
The initial condition $\chi_0(\lambda)$ is not
important for what follows (see below), so it is left unspecified.

In order to return to coordinate space, we need to perform the integral
\BQ\label{I}
I \equiv \int_C \frac{d\lambda}{2\pi i}\ {\rm e}^{\lambda r} 
\ {\rm e}^{\bar\alpha_s \tau
   \left\{2\psi(1)-\psi(\lambda)-\psi(1-\lambda)\right\}}\chi_0(\lambda)
 = \int_{C} \frac{d\lambda}{2\pi i} \,{\rm e}^{F(\lambda, r, \tau)}
\EQ
where $r\equiv \ln ({\rr^2 / \ell^2})$ is negative and large
in the range of interest.
(That is, in coordinate space, the second condition (\ref{INEQ}) is rewritten
as $(-r)\gg 1$.) 
The function (\ref{solchi}) has essential singularities at all the positive
integers $\la\ge 1$, so we can choose the contour as
$C=\{\lambda=a + i\nu, -\infty <\nu <\infty\}$, with $0<a<1$.

The integral (\ref{I}) will be evaluated 
in the saddle point approximation, which is a good approximation when $r$ and 
$\bar\alpha_s \tau$ are both large. 
(The corrections to it are suppressed by powers of
$1/r$ or $1/\bar\alpha_s \tau$.) To the same accuracy, we can ignore the
initial condition $\chi_0(\lambda)$, since its contribution to the function
$F(\lambda,r,\tau)$ in the exponent is not enhanced by either $r$ or 
$\bar\alpha_s \tau$. (We implicitly assume here that $\chi_0(\lambda)$
is not rapidly varying in the range of $\lambda$ of interest.)

We thus obtain: 
\BQ
I\, \simeq \,  {\rm e}^{F(\lambda_0)} \int_{-\infty}^\infty 
\frac{d\nu}{2\pi} \,{\rm e}^{- \frac12  \nu^2
F''(\lambda_0)} 
\,=\,{\rm e}^{F(\lambda_0)} \,\frac{1}{
{\sqrt{2\pi F''(\lambda_0)}}}\,,
 \label{Isaddle}
\EQ
where, from now on,
\BQ
F(\lambda)\equiv r \lambda + 
\bar\alpha_s\tau \left\{2\psi(1)-\psi(\lambda)-\psi(1-\lambda)\right\}
\equiv F(\lambda, r, \tau),
\EQ
and $\lambda_0$ satisfies the saddle point equation:
\BQ\label{SPEQ}
\frac{\del F(\lambda, r, \tau) }{\del \lambda}\biggl |_{\lambda_{0}}=0, \quad 
\lambda_0=\lambda_{0}(r,\tau).
\EQ
In eq.~(\ref{Isaddle}),
we have also included the contribution of the Gaussian fluctuations around
the saddle point. That is, we have taken the contour
$C=\{\lambda=\lambda_0 + i\nu, -\infty <\nu <\infty\}$, and we have expanded
in powers of $\nu$ to quadratic order before integrating over $\nu$. 

To visualize the solution to eq.~(\ref{SPEQ}), it is useful to keep in mind that
$2\psi(1)-\psi(\lambda)-\psi(1-\lambda)$ is a convex function of 
$\lambda$ with its minimum at $\lambda=1/2$ and 
simple poles at $\lambda=0$ and $\lambda=1$.  
A good approximation to this function in the range
$0<\lambda<1$ is given by \cite{Levin-Tuchin}
\BQ\label{app_F}
2\psi(1)-\psi(\lambda)-\psi(1-\lambda)
\approx {1\over \lambda}+ {1\over 1-\lambda}+ 4\ln 2 -4.
\EQ
Thus,  there is an unique saddle point $\lambda_0$ in the region 
$0<\lambda<1$, whose position
moves between 0 and 1 depending upon the value of the ratio $r/\bar\alpha_s\tau$.
There are three limiting cases of interest:

\bigskip 
\indent (A) When $r/\bar\alpha_s\tau$ is positive and large enough: 
the saddle point is close to $\lambda= 0$.

\indent (B) When $r/\bar\alpha_s\tau$ is small $\sim 0$:
   the saddle point is close to   $\lambda= 1/2$. 

\indent (C) When $r/\bar\alpha_s\tau$ is strongly negative: 
the saddle point is close to $\lambda= 1$.

\bigskip 

\noindent More
precise boundaries between these various cases will be specified below.

The first case is relevant when we consider the BFKL equation in  momentum 
space,
where $r\equiv \ln (\kk^2/\Lambda^2)$ is always positive in the range
of interest. Similarly, case (C) applies only to the coordinate space, 
where $r\equiv \ln (\rr^2/\ell^2)$ is {negative}. These two cases, 
(A) and (C), 
correspond to the double logarithmic approximation (DLA),
which describes the leading behaviour of the solution at large $\kk^2$
(or small $\rr^2$) for fixed, but large, $\tau$; this limit is common to the
BFKL and DGLAP equations.  Case (B), on the other hand,
applies to both the momentum space ($r>0$) and the coordinate space ($r<0$).
In standard analyses of the BFKL equation, this is the case which describes
the high energy limit ($\tau\to \infty$ at fixed $r$) \cite{AKM}. Here,
we shall never be truly in this limit, because of the condition 
(\ref{INEQ1}) which, strictly speaking, implies that 
$r/\bar\alpha_s\tau$ is never small~!
If case (B) is nevertheless relevant for us here, it is because the
true saddle point remains close to $\lambda= 1/2$ up to relatively
large values of $r/\bar\alpha_s\tau$, which, as we shall see,
leaves enough space for the condition (\ref{INEQ1}) to be satisfied.

\bigskip 

\noindent(A) Assuming that $\lambda_0\ll 1$, we find the 
saddle point
\BQ
\lambda_0 \simeq \sqrt{\frac{\bar\alpha_s \tau}{r}},
\EQ
which after insertion in eq.~(\ref{Isaddle}) and performing the Gaussian
integral there, leads to
\BQ\label{IDLA}
I\ \simeq\ {\rm e}^{2\sqrt{\bar\alpha_s \tau r}}\, 
\sqrt{\frac{\lambda_0^3}{4\pi \bar\alpha_s\tau}}.
\EQ
When $r=\ln (\kk^2/\Lambda^2)$, this gives the well-known ``DLA solution''
in the momentum space, which
coincides with the solution to the DGLAP equation in the double-log limit
\cite{DGLAP}.

The condition that $\lambda_0\ll 1$ should be more properly formulated as
$\lambda_0\ll 1/4$, since  $\lambda= 1/4$ is the middle point between
the limiting saddle points at $\lambda= 0$ and $\lambda= 1/2$.
This criterion gives the following 
range of validity for the DLA solution (\ref{IDLA}) in momentum space
\BQ\label{cond_momDLA}
16 \bar\alpha_s \tau \,\ll \,\ln (\kk^2/\Lambda^2).
\EQ

\bigskip 

\noindent (B) When $r/\bar\alpha_s\tau$ is small, the saddle point is very close to 
$\lambda=1/2$ :
\BQ\label{saddle_B}
\lambda_0 \simeq \frac12 - \delta\, , \qquad \ \delta\equiv \frac{r}{\beta\bar\alpha_s\tau},
\EQ
which gives:
\BQA
I&\simeq&  {\rm e}^{\omega\bar\alpha_s\tau} {\rm e}^{\frac12 r}
\exp \left\{ \frac{- r^2 }{2\beta\bar\alpha_s\tau} \right\}\,
\frac{1}{\sqrt{2\pi \beta\bar\alpha_s\tau} }\, ,
\label{linear_solution}
\EQA
where $ \beta\equiv (-2\psi''(1/2))=28\zeta(3)$ and
$\omega\equiv 2\psi(1)-2\psi(1/2)=4\ln 2$.

When $r=\ln(\kk^2/\Lambda^2)$, eq.~(\ref{linear_solution})
corresponds to the usual solution to the {\it momentum} BFKL equation.
On the other hand, when $r=\ln(\rr^2/\ell^2)$, it also gives the 
solution of the linearized BK equation in coordinate space:
\BQ\label{solution_cBFKL}
{\cal N}_\tau(\rr)\simeq \sqrt{\frac{\rr^2}{\ell^2}}
{{\rm e}^{\omega\bar\alpha_s\tau}\over\sqrt{2\pi \beta \bar\alpha_s \tau} } 
\exp \left\{ \frac{- \ln^2 \left({\rr^2/ \ell^2}\right) }{2\beta \bar\alpha_s\tau} 
\right\}.
\EQ  

The saddle
point in eq.~(\ref{saddle_B}) remains close to $1/2$ for all $r$ such that
 $\delta\ll 1/4$. This is realized  when (in momentum space, for definiteness)
\BQ\label{cond_BFKL}
\ln (\kk^2/\Lambda^2)
\,\ll \,8\bar\alpha_s\tau,
\EQ
which, as we shall soon discover, leaves a substantial window for genuine BFKL
behaviour (in the sense of eq.~(\ref{linear_solution})) above the saturation scale.

Since there is no overlapping between the conditions (\ref{cond_momDLA}) and
(\ref{cond_BFKL}), we conclude that in the intermediate range where $8\simle
\,r/\bar\alpha_s\tau\,\simle 16$, the saddle point is relatively close
to  $\lambda=1/4$. Indeed, by using the approximation (\ref{app_F}), one can
estimate that $\lambda_0=1/4$ for 
$r/\bar\alpha_s\tau\,\simeq 14$.

Eq.~(\ref{linear_solution}) is the solution that is usually
considered in applications of the BFKL equation at very high energy
\cite{TB-BFKL}. It shows an exponential growth with $\tau$
and ``infrared diffusion'' towards small $\kk$ momenta,
which would become problematic in the formal 
high energy limit $\tau\to \infty$ at fixed $r$ (which
is, of course, outside the kinematical range considered here;
recall eq.~(\ref{INEQ1}).)
But at sufficiently high energy, $Q_s^2(\tau) > Q^2$, and 
the BFKL equation gets supplanted by the BK equation 
(or the more general equations in Refs. \cite{B,W,PI}),
in which non-linear effects provide a natural solution
to the difficulties of eq.~(\ref{linear_solution})
alluded to above \cite{K,Levin-Tuchin,Motyka}.

\bigskip 

\noindent (C)
For $r/\bar\alpha_s\tau$  negative and large, the
saddle point is close to $\lambda=1$: 
\BQ\label{sp1}
\lambda_0\simeq 1-\delta_1\,,\qquad \delta_1\equiv 
\sqrt{\frac{\bar\alpha_s \tau}{(-r)}},
\EQ
and the integral is estimated as 
\BQ
I \ \simeq \ \exp \Big\{r+2\sqrt{\bar\alpha_s \tau (-r)}\Big\}\,
\sqrt{\frac{\delta^3_1}{4\pi \bar\alpha_s\tau}}.
\EQ
This saddle point is interesting only in coordinate space, 
where $r=\ln (\rr^2/\ell^2) < 0$, and gives the dominant behaviour 
${\cal N}_\tau(\rr)$ at short distances:
\BQ\label{BFKL-DLA}
{\cal N}_\tau(\rr)={\rr^2 \Lambda^2}\exp
\left\{2\sqrt{\bar\alpha_s \tau \ln \frac{1}{\rr^2\Lambda^2}}\right\}
\sqrt{\frac{\delta_1^3}{4\pi \bar\alpha_s\tau}}, \qquad 
\delta_1=\sqrt{\frac{\bar\alpha_s \tau}{\ln (1/\rr^2\Lambda^2)}}
\EQ
where we have taken $\ell^2=1/\Lambda^2$. As expected, this is the same
as the asymptotic solution of the DLA equation in coordinate 
space\footnote{The additional factor $\rr^2 \Lambda^2$ in front
of the exponential in eq.~(\ref{BFKL-DLA}) as compared to the DLA
solution (\ref{IDLA}) in momentum space comes up since, in coordinate
space, the DLA equation applies to ${\cal N}_\tau(\rr)/\rr^2$.}. 
The saddle point (\ref{sp1}) is 
close to 1 provided $\delta_1\ll 1/4$, which gives
\be
16 \bar\alpha_s \tau \,\ll \,\ln (1/\rr^2\Lambda^2),\ee
in complete agreement with the previous condition (\ref{cond_momDLA})
in momentum space.

It is convenient to summarize the previous results in terms of the
ratio 
\be\label{Rdef}
 R\,\equiv \,\frac{(-r)}{\bar\alpha_s\tau}\,=\,
\frac{1}{\bar\alpha_s\tau}\,\ln \frac{1}{\rr^2\Lambda^2}\,,\ee
($R$ is positive in the perturbative regime of interest)
which will play a special role in the next subsection. 
Our analysis shows that the saddle point solution to the BFKL equation
has the genuine BFKL  behaviour (\ref{linear_solution}) for $R\ll 8$, 
the DLA  behaviour (\ref{BFKL-DLA}) for $R\gg 16$, and some intermediate
behaviour in the window $8\simle R\simle 16$. These conclusions,
based on a pure BFKL analysis, are still to be combined with 
the constraint (\ref{INEQ1})  showing that we are indeed in the linear
regime. This requires to determine the coefficient $c$ in eq.~(\ref{INEQ1}),
which we shall do in the next subsection.

\subsection{Saturation scale and scaling from the BFKL equation}

As already explained, it is possible to determine the
saturation momentum from the solution ${\cal N}_\tau(\rr)$ 
to the BFKL equation (\ref{linearBK}) by using the saturation 
criterion:
\BQ\label{criterion}
{\cal N}_\tau(\rr=1/Q_s(\tau))\,=\,1.
\EQ
Strictly speaking, the number in the r.h.s. of this equation 
is not exactly 1 (cf. the discussion at the beginning of Sect. 3),
but this is irrelevant to the
accuracy of the present calculation. Indeed, replacing 1
by $\kappa < 1$ would modify the following results via subleading
terms, of relative order $(1/\bar\alpha_s \tau)\ln\kappa$.
Besides, such terms would affect
the overall normalization of the saturation scale (\ref{QSBFKL}),
that is, the value of the
reference scale $\Lambda^2$, which has not been fully
specified anyway.

In what follows, we shall rely on the saddle point approximation
in eqs.~(\ref{Isaddle})--(\ref{SPEQ})
to obtain an estimate
for $Q_s(\tau)$ \cite{AM2}, and then study the scaling
properties of the BFKL solution above the saturation scale.
For these purposes, it is crucial to notice from eq.~(\ref{SPEQ}), which is
rewritten as 
\BQ\label{SPE2}
\frac{\del}{\del \lambda}\Big(
\psi(\lambda)+\psi(1-\lambda)
\Big)\Big|_{\lambda_{0}}\,=\,\frac{r}{\bar\alpha_s \tau}\,\equiv \,-R\,,
\EQ
that the saddle point $\lambda_{0}$ is actually a function of only one variable $R$
(cf. eq.~(\ref{Rdef})) :
\BQ\label{scaling-lambda}
\lambda_0(r,\tau)\,=\,\lambda_0(R).
\EQ
If one estimates the Mellin integral just by the saddle point, one then obtains 
\BQ\label{IM}
{\cal N}_\tau(\rr)\,\simeq\, {\rm e}^{\bar\alpha_s \tau F(\lambda_0(R), R)},
\EQ
where, as compared to  eqs.~(\ref{Isaddle})--(\ref{SPEQ}), we have changed the
definition of the function $F(\lambda,r,\tau)$ by pulling out 
a factor $\bar\alpha_s \tau$. This is convenient since, when evaluated
at the saddle point,
the new function $F(\lambda_0(R), R)$ is only a function of $R$.
Also, we have neglected the (slowly varying)
factor coming from the Gaussian integral over the fluctuations.
This is correct up to  corrections
of relative order $\ln(\bar\alpha_s \tau)/\bar\alpha_s \tau$.
The effect of this factor will be illustrated in eq.~(\ref{QSDLA})
below.

The saturation criterion (\ref{criterion}) yields then the following
condition on $F$ :
\BQ\label{SAT-F}
F(\lambda_0(R_s), R_s)\,=\,0\qquad {\rm for}\qquad 
R_s\,=\,\frac{1}{\bar\alpha_s \tau}\ln \frac{Q_s^2(\tau)}{\Lambda^2}
\EQ
which is an equation for $R_s$, and ultimately for $Q_s(\tau)$.

This equation has two immediate and important consequences, which are
among the main results in this paper: (i) The
saturation momentum is increasing exponentially with $\tau$,
with the slope of the exponential uniquely fixed by the saddle point solution 
to the BFKL equation. (ii) The ``anomalous dimension'' which characterizes
the approach towards saturation (cf. eq.~(\ref{GS})),
is constant, i.e., independent of $\tau$, which then implies geometric scaling.

\bigskip

i) Indeed, the solution $R_s$ to eq.~(\ref{SAT-F}) is a pure number,
$R_s\equiv c$, and not a function of $\tau$ (as it could have been if the
function $F(\lambda,r,\tau)$ in eq.~(\ref{IM}) was a function
of two independent variables, $r$ and $\tau$, and not just of their ratio
$R$).
Together with the second equation (\ref{SAT-F}), this implies\footnote{If we
had included the Gaussian fluctuations around the saddle point, this
pure exponential would get multiplied by a slowly varying prefactor;
see eq.~(\ref{QSDLA}) for an example.}:
\BQ\label{QSBFKL}
Q_s^2(\tau)= \Lambda^2 \,{\rm e}^{c\bar\alpha_s \tau},
\EQ 
which is consistent with a previous result (\ref{saturation_scale_fixed})
based on scaling properties of the BK equation, but which here arises
from a (BFKL) solution which has no geometric scaling (e.g.,
eqs.~(\ref{linear_solution}) or (\ref{BFKL-DLA}) are not scaling functions
for generic values of $\rr$ and $\tau$). Moreover, the value of the
slope parameter $c$ is known here, since uniquely fixed by the
solution to eqs.~(\ref{SAT-F}) and (\ref{SPE2}).

In fact, in order to compute $R_s\equiv c$, there is no need 
to solve the saddle point equation (\ref{SPE2}) for arbitrary values of $R$ 
(although this could be done via numerical techniques). 
Rather, we notice that by combining eqs.~(\ref{SAT-F}) and (\ref{SPE2})
one can deduce an equation for the particular value $\lambda_s
\equiv \lambda_0(R_s)$ that the saddle point takes at saturation:
\be
\label{SPS}
-\lambda_{s}\frac{\del}{\del \lambda}\Big(
\psi(\lambda)+\psi(1-\lambda)
\Big)\Big|_{\lambda_{s}}\,=\,\lambda_{s}R_s\,=\,
2\psi(1)-\psi(\lambda_s)-\psi(1-\lambda_s).\ee
We have solved this numerically and obtained:
\be\label{lsrs}
\lambda_s\,=\,0.627549...,\qquad R_s\,\equiv\,c\,=\,4.88339....\ee
Note that this  $\lambda_s$ is not too far away from 1/2,
which is consistent with the fact that
the above value of $R_s$ is in the range where we expect a ``genuine''
BFKL behaviour, i.e., $R_s < 8$.
We shall return to this point after the
discussion of geometric scaling.

\bigskip

ii) Let us evaluate eq.~(\ref{IM})
for $R$ slightly above $R_s$, that is, for distances $\rr$ which,
while still being much shorter than the saturation length $1/Q_s(\tau)$,
are nevertheless close to it in logarithmic units. More precisely,
we shall require that (with $Q^2\equiv 1/\rr^2$) :
\be\label{window}
  0<\, R-R_s\,\ll\,R_s,\qquad {\rm or}\qquad
1<\,\ln \frac{Q^2}{Q_s^2(\tau)}\,\ll\,\ln \frac{Q_s^2(\tau)}{\Lambda^2}\,.\ee
The condition $R>R_s$ (i.e., $Q^2\gg Q_s^2(\tau)$) ensures that we stay
in the linear regime; in logarithmic units, this is effectively implemented
as $Q^2 > e Q^2_s(\tau)$, with $e=2.72...$. The condition $R-R_s\ll R_s$
allows us to study the approach of ${\cal N}_\tau(\rr)$ towards saturation
in a limited expansion in powers of $R-R_s$. To linear order in this
expansion, one has
\BQA\label{expF}
F(\lambda_0(R), R)&\simeq& 
F(\lambda_0(R_s),R_s)+\left.
\frac{d }{d R}F(\lambda_0(R), R)\right\vert_{R=R_s}
(R-R_s)
+\cdots\NN
&=&-\lambda_s (R-R_s) + \cdots,
\EQA
where we have used the saturation condition (\ref{SAT-F}), together with
the fact that $\lambda_0(R)$ is a solution of the
saddle point equation, cf. eq.~(\ref{SPEQ}), so that:
\BQA
\left.\frac{d}{d R}F(\lambda_0(R), R)\right\vert_{R=R_s}
\,=\,-\lambda_0(R_s) + 
\left.\frac{\del F}{\del \lambda}\right\vert_{\lambda=\lambda_0}
\left.\frac{\del \lambda_0}{\del R}\right\vert_{R=R_s}\,=\,-
\lambda_s.
\EQA
Therefore, above the saturation scale, 
the dipole-hadron scattering amplitude is approximated as follows
[recall that $\bar\alpha_s \tau(R-R_s)=\ln (1/\rr^2Q_s^2(\tau))$]
\BQA\label{GSBFKL}
{\cal N}_\tau(\rr)&\simeq&\kappa \,{\rm e}^{-\bar\alpha_s \tau\lambda_s(R-R_s)}\NN
&=& \kappa\left(\rr^2Q_s^2(\tau)\right)^{\lambda_s},
\EQA
where we have reintroduced the numerical factor $\kappa < 1$ (cf.
the discussion after eq.~(\ref{criterion})).

Eq.~(\ref{GSBFKL}) shows geometric scaling 
with the anomalous dimension\footnote{We
refer to $1-\lambda_s$ as an ``anomalous dimension'' since,
na\"{\i}vely, one could expect ${\cal N}_\tau(\rr)$ to vanish like
$\rr^2$ as $\rr\to 0$: this would follow from the fact that 
${\cal N}_\tau(0)=0$, together with analyticity near $\rr\to 0$.
In reality, this analyticity is broken
by logarithmic ultraviolet divergences in the formal expansion in
powers of $\rr^2$. These divergences correspond to the large
logarithms $\ln(Q^2/\Lambda^2)$ whose resummation by the DGLAP
(or DLA) equation leads to the non-analytic behaviour manifest
in eq.~(\ref{BFKL-DLA}). Besides, whatever are the analytic properties 
of the function ${\cal N}_\tau(\rr)$ near $\rr\to 0$ (i.e., at very high $Q^2$),
there is no reason why these properties should persist down to the
$Q^2=Q_s^2(\tau)$. Compare, in this respect, eqs.~(\ref{GSBFKL}) and
(\ref{BFKL-DLA}).}
$\gamma=1-\lambda_s\simeq 0.37$ determined  by the value of the saddle point 
$\lambda_s\equiv \lambda_0(R_s)$ at  saturation. 
We summarize here the two essential ingredients in the arguments leading
to this scaling: ({\it a}) the saturation condition which requires that
${\cal N}_\tau(\rr)\to 1$ when $\rr\to 1/Q_s(\tau)$; ({\it b}) the fact
that, at saturation, the saddle point $\lambda_0(R)$
becomes independent of $\tau$ (i.e., a pure number; cf. eq.~(\ref{lsrs})).
In turn, point ({\it b}) is the consequence of the ``scaling'' property 
(\ref{scaling-lambda}) of the saddle point, and ultimately reflects the
scale invariance of the BFKL equation (\ref{linearBK}).

The previous arguments also shed light on potential sources of 
scaling violation when going beyond the present approximations. 
For instance, we expect $\tau$--dependent corrections to the anomalous 
dimension due to the Gaussian fluctuations, the higher-order terms in the 
saddle point expansion, the initial condition $\chi_0(\la)$ in the Mellin function
(\ref{solchi}), etc. These corrections are suppressed by $1/\bar\alpha_s \tau$. 
Also, there are corrections to the functional form of
${\cal N}_\tau(\rr)$, coming from higher order terms in the expansion
(\ref{expF}) around the saturation scale. These  corrections are controlled
by the ratio $(R-R_s)/R_s =
\ln \frac{Q^2}{Q_s^2(\tau)}/\ln \frac{Q_s^2(\tau)}{\Lambda^2}$.
For instance, after also including the second order in this expansion,
eq.~(\ref{GSBFKL}) gets replaced by
\be\label{GSBFKLsec}
{\cal N}_\tau(\rr)\,\simeq\, \kappa\left(\rr^2Q_s^2(\tau)\right)^{\lambda_s}
\exp \left\{ -\frac{\lambda^\prime_s}{2 \bar\alpha_s\tau} \left(
\ln\frac{1}{\rr^2 Q_s^2(\tau)}\right)^2\right\} 
\ee
where $\lambda^\prime_s
\equiv (d\lambda_0(R)/dR)|_{R_s}$. Clearly, the exponential term
in this expression violates scaling.

The scaling behaviour (\ref{GSBFKL}) holds in a window in $Q^2$
specified by eq.~(\ref{window}), that is:
\be\label{wind}
1\,\,<\,\,\ln \frac{Q^2}{Q^2_s(\tau)}\equiv
\Big(\ln \frac{Q^2}{\Lambda^2}\,-\,c\bar\alpha_s\tau\Big)\,\,
\ll\,\,c\bar\alpha_s\tau.\ee
This requires that both $r=\ln ({Q^2}/{\Lambda^2})$ and
$c\bar\alpha_s\tau$ are large numbers, with $r$ larger
than $c\bar\alpha_s\tau$
(in agreement
with the original assumptions (\ref{INEQ}) and (\ref{INEQ1})), 
but not {\it much} larger: $r-c\bar\alpha_s\tau\ll c\bar\alpha_s\tau$.
In practice, this requires  $\ln ({Q^2}/{\Lambda^2})$
to be {\it deeply} inside the strip (note that $c=4.88...\simeq 5$)
\be
5\bar\alpha_s\tau\,<\,
\ln ({Q^2}/{\Lambda^2})\,<\,10\bar\alpha_s\tau\,,\ee
which is indeed a large window when $5\bar\alpha_s\tau\gg 1$.
Together with eq.~(\ref{cond_BFKL}), this suggests that
the window for extended scaling is almost entirely located
in the kinematical range controlled by the BFKL saddle point 
(\ref{saddle_B}).

To verify that, let us compute directly the predictions of this
saddle point and the associated solution (\ref{solution_cBFKL}) for
the saturation scale and geometric scaling. The ``BFKL saturation scale'' 
is defined by imposing
the condition (\ref{criterion}) directly on eq.~(\ref{solution_cBFKL}):
\BQ\label{SAT-BFKL}
\sqrt{\frac{\Lambda^2}{Q^2_s}}\,
{{\rm e}^{\omega\bar\alpha_s\tau}}\,
\exp \left\{ \frac{- \ln^2 \left(Q^2_s/\Lambda^2\right)} {2\beta \bar\alpha_s\tau} 
\right\}\,=\,1.
\EQ  
As expected, this amounts to an equation for $R_s$, the value
of the variable $R$ at saturation (compare to eq.~(\ref{SAT-F})).
In this case, this is a second-order equation:
\be R_s^2 + \beta R_s -2\beta\omega\,=\,0,\ee
with the  positive solution:
\be\label{RsBFKL}
R_s\Big|_{\rm BFKL}\,=\,\frac{1}{2}\,\Big(-\beta+\sqrt{\beta(\beta+8\omega)}
\Big)\,= \,4.8473....\ee
This is the value of the slope parameter $c$ predicted by the standard
BFKL solution, and is numerically very close to that in 
eq.~(\ref{lsrs}). By using (\ref{RsBFKL}) and (\ref{saddle_B}),
one can compute the BFKL saddle point at saturation $\lambda_s$ (or the
anomalous dimension $\gamma=1-\lambda_s$):
\be\label{lamBFKL}
\lambda_s\Big|_{\rm BFKL} \simeq \frac12 + \frac{R_s}{\beta}\,=\,
0.644...,\ee
which is indeed very close to the true saddle point (\ref{lsrs}).
This shows that for all practical purposes one can use the explicit
BFKL solution (\ref{solution_cBFKL}) for any $Q^2=1/\rr^2$ in the
window for extended scaling.
In fact, without any approximation, eq.~(\ref{solution_cBFKL}) can be cast
in the form of the ``second-order expansion'' in eq.~(\ref{GSBFKLsec}),
with $\lambda_s$ given 
by eq.~(\ref{lamBFKL}), and $\lambda^\prime_s=1/\beta \simeq 1/33.67$.
This latter is a 
rather small number, showing that the violations of scaling due
to the exponential term are only tiny. That is, in its whole domain of applicability,
the ``genuine'' BFKL solution (\ref{solution_cBFKL}) is almost an 
exact scaling solution.

For completeness, and also for comparison with previous analytic
studies in the literature which have used the DLA approximation
\cite{Levin-Tuchin,SAT,Motyka}, let us finally
evaluate the scaling predictions of the DLA saddle point
(\ref{sp1}). The saturation condition (\ref{criterion}) applied to
eq.~(\ref{BFKL-DLA}) yields:
\BQ\label{QSDLA}
Q_s^2(\tau)\Big|_{\rm DLA}\simeq\, \Lambda^2 \,\frac{{\rm e}^{4\bar\alpha_s \tau}}
{{32\pi\bar\alpha_s\tau}},
\EQ 
where the slowly varying factor multiplying the exponential 
is the effect of the Gaussian fluctuations around the saddle point.
Eq.~(\ref{QSDLA}) implies:
\be
R_s\Big|_{\rm DLA}\,\simeq\,4 - \frac{\ln({32\pi\bar\alpha_s\tau})}{\bar\alpha_s\tau}
\,\simeq\,4\,,\ee
which is slowly dependent upon $\tau$, because of the contribution of the
fluctuations, but to leading order takes the constant value
$c_{\rm DLA}=4$. This is in agreement with previous studies \cite{Levin-Tuchin,SAT,Motyka}.
If inserted in eq.~(\ref{sp1}), it gives the 
DLA anomalous dimension $\gamma=1-\lambda_s$
\cite{Levin-Tuchin}: 
\BQ\label{DLA-SP}
\gamma\Big|_{\rm DLA}\,=\,\sqrt{\frac{1}{R_s}}\,\simeq\,\frac12
+ \frac{\ln({32\pi\bar\alpha_s\tau})}{16 \bar\alpha_s\tau}\,\simeq\,
\frac12\,.
\EQ 
This is slightly larger 
than for the true saddle point, or the BFKL saddle point (\ref{lamBFKL}).
The DLA window for extended geometric scaling is given by eq.~(\ref{wind})
with $c=4$. 

Thus, at a first sight, the DLA predictions look rather similar
to those of the BFKL approximation, both for the value of the
saturation scale (which fixes also the scaling window),
and for the anomalous dimension. However, one should keep in mind
that these predictions are not consistent with the validity region of
DLA: the  saddle point (\ref{DLA-SP})
not only is not close to 1, but it even takes the typical BFKL
value. Besides, the corresponding scaling window is truly within
the kinematical range for BFKL.

We finally mention that numerical studies of the BK equation
\cite{Levin-Tuchin,AB01,Motyka} (with a fixed coupling) have 
found that the saturation scale is indeed increasing exponentially
with $\tau$, with a slope parameter $c\approx 4.1$ which is 
intermediate between the DLA and BFKL predictions obtained in this
section.

\subsection{The running coupling  case: $\alpha_s(Q_s^2(\tau))$}

Except for the modified $\tau$--dependence of the saturation scale,
which changes according to the general expectation in 
eq.~(\ref{saturation_scale_run}), all the previous discussion of the
solutions to the BFKL equation and their consequences for extended scaling
goes almost unchanged to the case of a running coupling 
in which the scale for running is set by $Q_s(\tau)$. In this subsection,
we shall only indicate the few steps which involve non-trivial differences.

The extra $\tau$ dependence of the coupling $\alpha_s(Q_s^2(\tau))$
commutes with the Mellin transform (\ref{Mellin}), which is defined only 
in terms of the transverse coordinates. Thus, the previous
equation for $\chi_\tau(\lambda)$ is simply replaced by 
\BQA
\frac{\del}{\del \tau} \chi_\tau(\lambda)
&=&\frac{b_0}{\ln (Q_s^2(\tau)/\Lambda_{\rm QCD}^2)}\,   
\left\{2\psi(1)-\psi(\lambda)-\psi(1-\lambda)\right\} \chi_\tau(\lambda).
\EQA
So far, the saturation scale is not known, but it can be absorbed
into a redefinition of the ``time'' variable $\tau$ :
\BQ\label{newtau}
\ln (Q_s^2(\tau)/\Lambda_{\rm QCD}^2)\, \frac{\del}{\del \tau}\,\equiv\,
\frac{\del}{\del \tilde\tau},\qquad {\rm or}\qquad 
\tilde \tau \,\equiv\, \int^\tau_0 d\tau' \frac{1}{f(\tau')}\,,
\EQ
where we have written $Q_s^2(\tau)=\Lambda_{\rm QCD}^2 {\rm e}^{f(\tau)}$. 
Then, the equation is easily solved: 
\BQ
\chi_\tau(\lambda)= {\rm e}^{b_0 \tilde \tau
\left\{2\psi(1)-\psi(\lambda)-\psi(1-\lambda)\right\} },
\EQ
which leads us to the following Mellin representation for
the solution to the BFKL equation with running coupling:
\BQ\label{Solution}
{\cal N}_\tau(\rr)=\int_{C} \frac{d\lambda}{2\pi i} \, 
{\rm e}^{F(\lambda, r, \tilde \tau)}\quad {\rm with} \quad 
F(\lambda, r, \tilde\tau) = r \lambda + 
b_0 \tilde \tau
\left\{2\psi(1)-\psi(\lambda)-\psi(1-\lambda)\right\}
\EQ
where $r=\ln ({\rr^2 / \ell^2})=\ln (\rr^2\Lambda^2) $ as before.
As anticipated, this has the same structure as in the fixed 
coupling case (cf. eq~(\ref{I})). Thus, all
the results in Sects. 3.1 and 3.2 can 
be immediately translated to the case of a running coupling by simply replacing
\be
\bar\alpha_s\,\rightarrow \,b_0,\qquad {\rm and}\qquad \tau\,\rightarrow \,
\tilde \tau,\ee
in the corresponding formulae. In particular, 
by the same arguments as before, the saddle point $\lambda_0$ is a function 
of the ratio $R$ alone, with $R\equiv (-r)/b_0 \tilde \tau$. 
The value $\lambda_s=\lambda_0(R_s)$ of the saddle point at saturation is 
again determined by eq.~(\ref{SPS}), so that $\lambda_s$ and
$c\equiv R_s$ take the same values as before, cf. eq.~(\ref{lsrs}). 
Thus, all the previous results on extended scaling --- the value of the anomalous 
dimension $\gamma=1-\lambda_s$ in eq.~(\ref{GSBFKL}),
and the momentum range (\ref{window}) in which the scaling holds --- 
remain unchanged, except for the expression of the saturation scale
which enters these results.

To determine this scale, we use
$c=R_s=(-r_s)/b_0 \tilde \tau$ with 
$(-r_s)=\ln Q_s^2(\tau)/\Lambda_{\rm QCD}^2=f(\tau)$, together
with eq~(\ref{newtau}), to derive an equation for $f(\tau)$:
\BQ
{ 1\over c} f(\tau)\,= \,b_0 \int^\tau_0 d\tau' \frac{1}{f(\tau')}.
\EQ
This has the solution
$f(\tau)=\sqrt{2b_0 c (\tau+\tau_0)}$
(and therefore $\tilde \tau =\sqrt{2(\tau+\tau_0)/b_0 c}\,$).
Therefore, the saturation scale is determined as 
\BQ\label{satu}
Q_s^2(\tau)=\Lambda_{\rm QCD}^2 {\rm e}^{\sqrt{2b_0 c (\tau+\tau_0)}},
\EQ
where $\tau_0$ is arbitrary and $c$ is given by eq.~(\ref{lsrs}). 
This has the same functional form as eq.~(\ref{saturation_scale_run}) that
has been obtained from the BK equation with a scaling Ansatz.
In particular, this confirms the previous arguments in Sect. 2 that the
coefficient $c$ which enters the exponent of the saturation scale
should be the same for fixed coupling and running coupling.

Needless to say, the BFKL and DLA predictions for the anomalous dimension,
eqs.~(\ref{lamBFKL}) and respectively (\ref{DLA-SP}), remain unchanged.

\section{Summary and discussion}

We have shown that the geometric scaling predicted at low momenta 
$Q^2 \simle Q_s^2$ by the Colour Glass Condensate and 
phenomenological saturation models is preserved by the BFKL evolution
equation up to relatively large $Q^2$ momenta, within the range
$1 \simle \ln(Q^2/Q_s^2) \ll \ln(Q_s^2/\Lambda^2_{\rm QCD})$.
By matching the solution to the BFKL equation onto the saturation
condition at $Q^2 \sim Q_s^2$, we have determined the dependence
of the saturation scale $Q_s$ upon the rapidity $\tau=\ln(1/x)$,
and the anomalous dimension of the distribution function near
saturation.

The matching has been performed for the dipole scattering amplitude,
which enters linearly the structure function $F_2(x,Q^2)$,
and can be also related to the gluon distribution.
We have found the same kinematical window for extended scaling 
for both fixed and running couplings (with the scale
for running set by $Q_s(\tau)$), although
the $\tau$--dependence of the saturation momentum turns out to
be different in the two cases. We have also found that, formally,
the double logarithmic approximation to the DGLAP equation 
leads to qualitatively, and even quantitatively,
similar predictions for the extended scaling, but these results are 
inconsistent with the validity range of this approximation.

We have shown that the functional form of the $\tau$--dependence 
of the saturation scale $Q_s$ which follows
from the BFKL equation is consistent with the general scaling
properties of the BK equation.

It would be extremely interesting to compare these results
with the $F_2$ data in deep inelastic scattering, for
which geometric scaling has been originally identified \cite{geometric},
and to explore possible implications for particle production in 
heavy-ion experiments, where a phenomenological scaling law
has been recently reported \cite{MSB01}.
In particular, we expect these
results to have consequences for the analysis of the
multiplicity distributions of produced particles in heavy ion 
collisions at RHIC \cite{Dima}.  Our analysis shows that
while $F_2$ has geometric scaling, the gluon distribution function,
due to an extra factor of $1/\alpha_s$ in
 eqs.~(\ref{G-def})--(\ref{glue}), has logarithmic scaling violations.  
This is consistent with scaling violations
which are seen in the RHIC data \cite{Dima}.

While the qualitative phenomenon of the existence of extended
geometric scaling up to momenta  $Q^2$ of order $100~{\rm GeV}^2$
is certainly consistent with the analysis of deep inelastic scattering
by Sta\'sto, Golec-Biernat and Kwieci\'nski \cite{geometric},
it is at the same time clear that the saturation scale emerging
from our analysis (for either fixed or running coupling) is
too rapidly increasing with $\tau=\ln(1/x)$ to give a good
description of the data for $F_2$.

Consider fixed coupling first. Although eq.~(\ref{QSBFKL})
shows a power law increase, $Q^2_s(x)=Q^2_0x^{-\lambda}$, in
agreement with the parametrization used in Ref. \cite{geometric},
the actual 
value of the parameter $\lambda$ predicted by the BFKL equation,
namely $\lambda \simeq (4 - 5)\alpha_s N_c/\pi$,
is sensibly larger (for $N_c=3$ and realistic
values of $\alpha_s$) than the value $\lambda = 0.3 - 0.4$
extracted from the fit to the data \cite{geometric}.

As for the corresponding prediction in the case of a running coupling, 
eq.~(\ref{satu}), this is less rapidly increasing at very large $\tau$
: $Q^2_s(\tau)=\Lambda_{\rm QCD}^2\exp\{\sqrt{C(\tau + \tau_0)}\}$.
But since the number $C=2b_0 c$ is relatively large
(of  order 10; cf. eqs.~(\ref{running}) and (\ref{lsrs})),
this too fails to reproduce the data,
unless the parameter $\tau_0$ is mysteriously high.

The reasons for such a failure can be several. First, there are
the many approximations that we have performed in order to obtain
analytic solutions to the BFKL equation. Note that we have preserved
just leading order terms in the {\it exponents}, and that the subleading
terms are truly small only if the inequalities (\ref{INEQ}) are 
strictly satisfied. These are strong inequalities on logarithms,
which may not be well fulfilled for realistic values of
$x$ and $Q^2$. Besides,
even small subleading terms may give a substantial effect once
exponentiated. This is illustrated by eqs.~(\ref{QSDLA})--(\ref{DLA-SP})
which, in addition to the lowest order terms, include also
the corrections due to the Gaussian fluctuations of the saddle point.
As obvious on eq.~(\ref{QSDLA}), these ``subleading'' terms
may drastically change the actual value of the saturation momentum.
Still, the fact that the slope parameter $c\approx 4 - 5$ that we 
have found is rather close to that  obtained 
via numerical studies of the BK equation \cite{AB01,Motyka} makes us
confident about our control of the approximation scheme
for the exponent.

The last argument also suggests that, independent of further
approximations, there is a true
discrepancy between the  exponent $\lambda \simeq 4\alpha_s$ 
predicted by the BK equation and its
phenomenological value $\lambda = 0.3 - 0.4$ \cite{geometric,GBW99}.
Recall that the BK equation is strictly valid only
in the large $N_c$ limit, and to lowest order in $\alpha_s$
(within a leading-log approximation scheme).
For finite $N_c$, there is no simple
closed non-linear equation, just an infinite hierarchy of
coupled equations \cite{B},
which is equivalent to a functional Fokker-Planck equation \cite{W,PI}. 
This functional equation can be studied via numerical techniques
(in particular, on the basis of the associated Langevin
equation \cite{W}, or on the path integral representation \cite{BIW}),
and it would be very interesting to estimate the finite $N_c$ corrections.
Moreover, higher-order corrections in $\alpha_s$, which so far
have not been included in the non-linear evolution
equations (see however \cite{BB01}), but which are now available
for the BFKL kernel \cite{NLBFKL}, may be responsible for an
effective decrease in the slope  parameter $\lambda$ with respect to
its lowest order BFKL value. 

Note finally that the uncertainty on the $\tau$--dependence
of the saturation scale should not affect our prediction 
for the kinematical window in which one expects extended scaling.
Indeed, as explained in relation with eq.~(\ref{A}), this
prediction relies just on a limited expansion 
around
$Q^2=Q^2_s$, together with the scale-invariance of the linear
evolution equation at hand (in our case, BFKL). This argument
predicts an upper limit $Q^2_{max}\sim
Q_s^4/\Lambda_{\rm QCD}^2$ up to which geometric scaling should
be expected. To estimate this upper limit,
we shall use phenomenologically reasonable
values for $Q_s$ \cite{GBW99,geometric}, 
and not the theoretical predictions of our
analysis (which cannot give the absolute value of $Q_s$ anyway,
just its $\tau$--dependence). This gives 
$Q^2_{max}\sim  100~ {\rm GeV}^2$ for $Q_s \sim 1~{\rm GeV}$,
and $Q^2_{max}\sim 400~ {\rm GeV}^2$ for $Q_s \sim 2~{\rm GeV}$,
values which are both of the right order of magnitude to agree with
the phenomenological analysis in Ref. \cite{geometric}.

\section*{Acknowledgments}

We gratefully acknowledge conversation with Alfons Capella,
Miklos Gyulassy, Xin-Nian Wang, and, especially, Dima Kharzeev
whose provocative comments
forced us into thinking about these problems.  
This manuscript has been authorized under Contract No. DE-AC02-98H10886 with
the U. S. Department of Energy.

\section*{Note added}

Very recently, after this work was essentially completed, a paper
has been released by Kwieci\'nski and Sta\'sto \cite{KS02} which addresses
the issue of geometric scaling at high $Q^2$ in the framework of
the DGLAP evolution equation (or its double-log approximation).
The conclusions in this paper are however different from 
ours: The authors of Ref.  \cite{KS02} have not
recognized the existence of a window for geometric scaling above
$Q_s$, but rather concluded that scaling violations should be 
expected for any $Q^2>Q^2_s(x)$, even at very small values of $x$.

\end{document}